\documentclass[a4paper]{article} 
\usepackage[T1]{fontenc} 
\usepackage[utf8]{inputenc} 
\usepackage{ISMIRlbd, cite, times, amsmath, url}
\usepackage{hyperref}
\usepackage{url}
\usepackage{float}
\usepackage{graphicx}
\usepackage{booktabs}
\usepackage{scrextend}
\usepackage{color}
\usepackage{subfig}

\title{musicnn: pre-trained convolutional neural networks for music audio tagging}

\twoauthors
  {Jordi Pons} {Music Technology Group\\Universitat Pompeu Fabra\\ {\tt jordi.pons@upf.edu}}
  {Xavier Serra} {Music Technology Group\\Universitat Pompeu Fabra\\ {\tt xavier.serra@upf.edu}}

\begin{document}

\maketitle

\begin{abstract}

Pronounced as "musician", the \textit{musicnn} library contains a set of pre-trained musically motivated convolutional neural networks~\cite{pons2017end,pons2016experimenting} for music audio tagging:

\begin{center}
	\url{https://github.com/jordipons/musicnn}
\end{center}

This repository also includes some pre-trained vgg-like baselines~\cite{choi2016automatic}. These models can be used as out-of-the box music audio taggers, as music feature extractors, or as pre-trained models for transfer learning.

These models are trained with two different datasets: the MagnaTagATune dataset (the MTT of ~19k training songs~\cite{law2009evaluation})\footnote{The MagnaTagATune 50-tags vocabulary: \textit{guitar, classical, slow, techno, strings, drums, electronic, rock, fast, piano, ambient, beat, violin, vocal, synth, female, indian, opera, male, singing, vocals, no vocals, harpsichord, loud, quiet, flute, woman, male vocal, no vocal, pop, soft, sitar, solo, man, classic, choir, voice, new age, dance, male voice, female vocal, beats, harp, cello, no voice, weird, country, metal, female voice, choral.}} and the Million Song Dataset (the MSD of ~200k training songs~\cite{bertin2011million})\footnote{The Million Song Dataset 50-tags vocabulary: \textit{rock, pop, alternative, indie, electronic, female vocalists, dance, 00s, alternative rock, jazz, beautiful, metal, chillout, male vocalists, classic rock, soul, indie rock, mellow, electronica, 80s, folk, 90s, chill, instrumental, punk, oldies, blues, hard rock, ambient, acoustic, experimental, female vocalist, guitar, hip-hop, 70s, party, country, easy listening, sexy, catchy, funk, electro, heavy metal, progressive rock, 60s, rnb, indie pop, sad, house, happy}.}. 

\textbf{Which pre-trained models are available?} Although the main focus of the library is to release pre-trained musically motivated convolutional neural networks, we also provide several vgg-like models\footnote{\label{mark2}An in-depth depiction of \textit{vgg} architecture and its feature-maps is accessible online via a Jupyter Notebook: \url{https://github.com/jordipons/musicnn/blob/master/vgg_example.ipynb}} (as baselines for comparison). A high-level depiction of the \textit{musicnn} architecture\footnote{\label{mark1}An in-depth depiction of the \textit{musicnn} architecture (musically motivated CNN) and its feature-maps is accessible online via a Jupyter Notebook: \url{https://github.com/jordipons/musicnn/blob/master/musicnn_example.ipynb}} is depicted in the following figure:
\vspace{2mm}
\begin{figure}[H]
	\centerline{
		\includegraphics[width=0.75\columnwidth]{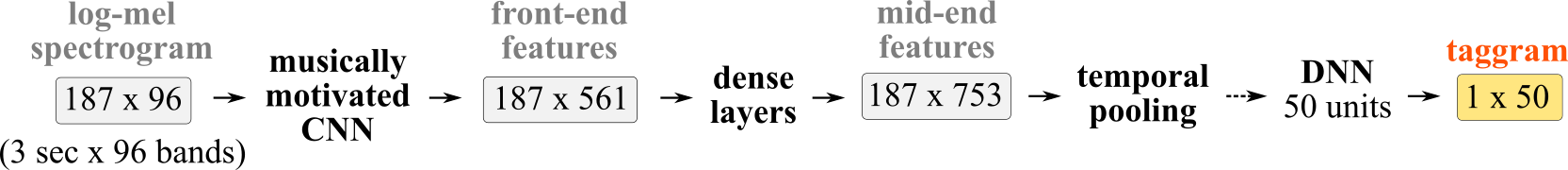}}
	\caption{The \textit{musicnn} architecture: a musically motivated convolutional neural network~\cite{pons2016experimenting,pons2017end}.}
	\label{fig:example}
\end{figure}
\vspace{-3mm}
The following models are available: \verb|MTT_musicnn|, \verb|MSD_musicnn|, \verb|MSD_musicnn_big|, \verb|MTT_vgg|, and \verb|MSD_vgg|. The \verb|MTT| models are trained with the MagnaTagATune dataset, and the \verb|MSD| models are trained with the Million Song Dataset. Given that the Million Song Dataset contains more training data, we also provide a larger \textit{musicnn} model: \verb|MSD_musicnn_big|. Architectural details are accessible online\footref{mark1}\footref{mark2}.

\textbf{What's the \textit{musicnn} library for? Out-of-the-box music audio tagging.} 
From within python, one can estimate the top-10 tags by simply running:
\vspace{-2mm}
\begin{verbatim}
from musicnn.tagger import top_tags
top_tags('music_file.mp3', model='MTT_musicnn', topN=10)
\end{verbatim}
\phantom{}
\vspace{-9mm}
From the command-line, one can also print the top-N tags on the screen (top) or save them to a file (bottom):
\vspace{-7mm}
\begin{verbatim}
python -m musicnn.tagger music.au --model 'MTT_musicnn' --topN 10 --print
python -m musicnn.tagger audio.wav -m 'MTT_vgg' --topN 5 --save out.tags
\end{verbatim}
\textbf{What's the \textit{musicnn} library for? Music feature extraction.} 
Out of the extractor, see the example below, one gets the output of the model (the taggram and its associated tags) and all the intermediate representations of it (we refer to those as features). The features are packed in a dictionary and, for the \textit{musicnn} models, you can extract \verb|timbral|, \verb|temporal|, \verb|cnn1|, \verb|cnn2|, \verb|cnn3|, \verb|mean_pool|, \verb|max_pool|, and \verb|penultimate| features\footref{mark1}. For the \textit{vgg} models, you can extract \verb|pool1|, \verb|pool2|, \verb|pool3|, \verb|pool4|, and \verb|pool5| features\footref{mark2}. 
\vspace{-2mm}
\begin{verbatim}
from musicnn.extractor import extractor
output = extractor(file, model='MTT_musicnn', extract_features=True)
taggram, tags, features = output
\end{verbatim}

\textbf{What's the \textit{musicnn} library for? Transfer learning.} Our pre-trained deep learning models can be fine-tuned, together with an output neural-network that acts as a classifier, to perform any other music~task. To assess the utility of our embeddings, we build SVM classifiers on top of several pre-trained models that act as music feature extractors.
The tools used to run this simple transfer learning experiment are accessible online:

\begin{center}
		\url{https://github.com/jordipons/sklearn-audio-transfer-learning}
\end{center}

We report accuracy results on the test set of the GTZAN (fault-filtered) dataset, and our processing pipeline consists of ``feature extraction'' + 128 PCA + SVM. The feature extraction can be based on VGGish audioset features (77.58\%~accuracy), OpenL3 audioset features (74.65\%~accuracy), \verb|MTT_musicnn| features (71.37\%~accuracy), \verb|MTT_vgg| features (72.75\%~accuracy), or \verb|MSD_musicnn| features (77.24\%~accuracy). Note that our MSD pre-trained models outperform the MTT ones. Besides, the \verb|MSD_musicnn| achieves similar results than the VGGish audioset features (that are trained with a much larger dataset: 2M audios).



\textbf{How did you train \textit{musicnn} models?} The code we employed to train the models above is also accessible:

\begin{center}
	\url{https://github.com/jordipons/musicnn-training}
\end{center}

These models achieve state-of-the-art performance on the MagnaTagATune dataset: \verb|MTT_musicnn| \mbox{(90.69 ROC-AUC / 38.44 PR-AUC)} and \verb|MTT_vgg| \mbox{(90.26 ROC-AUC / 38.19 PR-AUC)}. But also for the Million Song Dataset: 
\verb|MSD_musicnn| \mbox{(88.01 ROC-AUC / 28.90 PR-AUC)}, 
\verb|MSD_musicnn_big| \mbox{(88.41 ROC-AUC / 30.02 PR-AUC)} and
\verb|MSD_vgg| \mbox{(87.67 ROC-AUC / 28.19 PR-AUC)}.


But the \verb|musicnn-training| framework also allows to implement other models. For example, a similar architecture than \textit{musicnn} but with an attention-based output layer (instead of the temporal pooling layer) can achieve 90.77 ROC-AUC / 38.61 PR-AUC on the MagnaTagATune dataset --- and 88.81 ROC-AUC / 31.51~PR-AUC on the Million Song Dataset. You can find further details about this new architecture online.

%
%
%
%
%
%
%

\end{abstract}


\begin{acknowledgments}
This work was partially supported by the Maria de MaeztuUnits of Excellence Programme (MDM-2015-0502) --- and we are grateful for the GPUs donated by NVidia.
\end{acknowledgments}

\bibliography{ISMIRtemplate}

\end{document}